\documentclass{article}
\usepackage{amsmath, amssymb, amsfonts}
\title{On the conversion of rest energy in horizon energy} 
\author{Hristu Culetu, \\Ovidius University, Dept.of Physics, \\B-dul Mamaia 124, 900527 Constanta, Romania, \\e-mail : hculetu@yahoo.com}

\begin{document}
\numberwithin{equation}{section}
\pagenumbering{arabic}
\maketitle
\newcommand{\fv}{\boldsymbol{f}}
\newcommand{\tv}{\boldsymbol{t}}
\newcommand{\gv}{\boldsymbol{g}}
\newcommand{\OV}{\boldsymbol{O}}
\newcommand{\wv}{\boldsymbol{w}}
\newcommand{\WV}{\boldsymbol{W}}
\newcommand{\NV}{\boldsymbol{N}}
\newcommand{\hv}{\boldsymbol{h}}
\newcommand{\yv}{\boldsymbol{y}}
\newcommand{\RE}{\textrm{Re}}
\newcommand{\IM}{\textrm{Im}}
\newcommand{\rot}{\textrm{rot}}
\newcommand{\dv}{\boldsymbol{d}}
\newcommand{\grad}{\textrm{grad}}
\newcommand{\Tr}{\textrm{Tr}}
\newcommand{\ua}{\uparrow}
\newcommand{\da}{\downarrow}
\newcommand{\ct}{\textrm{const}}
\newcommand{\xv}{\boldsymbol{x}}
\newcommand{\mv}{\boldsymbol{m}}
\newcommand{\rv}{\boldsymbol{r}}
\newcommand{\kv}{\boldsymbol{k}}
\newcommand{\VE}{\boldsymbol{V}}
\newcommand{\sv}{\boldsymbol{s}}
\newcommand{\RV}{\boldsymbol{R}}
\newcommand{\pv}{\boldsymbol{p}}
\newcommand{\PV}{\boldsymbol{P}}
\newcommand{\EV}{\boldsymbol{E}}
\newcommand{\DV}{\boldsymbol{D}}
\newcommand{\BV}{\boldsymbol{B}}
\newcommand{\HV}{\boldsymbol{H}}
\newcommand{\MV}{\boldsymbol{M}}
\newcommand{\be}{\begin{equation}}
\newcommand{\ee}{\end{equation}}
\newcommand{\ba}{\begin{eqnarray}}
\newcommand{\ea}{\end{eqnarray}}
\newcommand{\bq}{\begin{eqnarray*}}
\newcommand{\eq}{\end{eqnarray*}}
\newcommand{\pa}{\partial}
\newcommand{\f}{\frac}
\newcommand{\FV}{\boldsymbol{F}}
\newcommand{\ve}{\boldsymbol{v}}
\newcommand{\AV}{\boldsymbol{A}}
\newcommand{\jv}{\boldsymbol{j}}
\newcommand{\LV}{\boldsymbol{L}}
\newcommand{\SV}{\boldsymbol{S}}
\newcommand{\av}{\boldsymbol{a}}
\newcommand{\qv}{\boldsymbol{q}}
\newcommand{\QV}{\boldsymbol{Q}}
\newcommand{\ev}{\boldsymbol{e}}
\newcommand{\uv}{\boldsymbol{u}}
\newcommand{\KV}{\boldsymbol{K}}
\newcommand{\ro}{\boldsymbol{\rho}}
\newcommand{\si}{\boldsymbol{\sigma}}
\newcommand{\thv}{\boldsymbol{\theta}}
\newcommand{\bv}{\boldsymbol{b}}
\newcommand{\JV}{\boldsymbol{J}}
\newcommand{\nv}{\boldsymbol{n}}
\newcommand{\lv}{\boldsymbol{l}}
\newcommand{\om}{\boldsymbol{\omega}}
\newcommand{\Om}{\boldsymbol{\Omega}}
\newcommand{\Piv}{\boldsymbol{\Pi}}
\newcommand{\UV}{\boldsymbol{U}}
\newcommand{\iv}{\boldsymbol{i}}
\newcommand{\nuv}{\boldsymbol{\nu}}
\newcommand{\muv}{\boldsymbol{\mu}}
\newcommand{\lm}{\boldsymbol{\lambda}}
\newcommand{\Lm}{\boldsymbol{\Lambda}}
\newcommand{\opsi}{\overline{\psi}}
\renewcommand{\tan}{\textrm{tg}}
\renewcommand{\cot}{\textrm{ctg}}
\renewcommand{\sinh}{\textrm{sh}}
\renewcommand{\cosh}{\textrm{ch}}
\renewcommand{\tanh}{\textrm{th}}
\renewcommand{\coth}{\textrm{cth}}

\begin{abstract}
 It is shown that the Verlinde formula for the entropy variation of a holographic screen is a consequence of the conversion of the particle energy in horizon energy. The special role played by the particular displacement $\Delta x = c^{2}/a$ is emphasized, $a$ being the particle acceleration. Using the Heisenberg Principle we show that the energy on the causal horizon (viewed as a holographic screen) of an inertial observer is proportional to its radius, as for a black hole.\\
\textbf{PACS} : 04.90.+e, 05.90.+m, 04.70.Dy.\\
\textbf{Keywords}: horizon energy, information erasing, Unruh effect, surface gravity. 
\end{abstract}

\maketitle

 \section{Introduction} 
  The discovery of the Unruh effect and the Hawking radiation \cite{UH} strengthened the reasons to look for a deep connection between gravity, quantum and thermodynamics.
 
 Since 1994, 't Hooft conjectured that all the degrees of freedom of an isolated system are located on a two surface surrounding the region (the Holographic Principle - HP), with the entropy proportional to its area in Planck units \cite{HB}. Padmanabhan \cite{TP} has derived Newton's law of gravity from the Unruh formula for the temperature of the thermal bath, the HP and the Equipartition law. His idea was further developed by Verlinde \cite{EV} who obtained the Newton law from thermodynamics by means of a new formula for the entropy variation of a holographic screen :
 \begin{equation}
 \Delta S = 2 \pi k_{B} \frac{mc}{\hbar} \Delta x
 \label{1}
 \end{equation}
 where $\Delta x$ is the distance between the screen and some test particle of mass $m$, $k_{B}$ is the Boltzmann constant, $c$ is the speed of light and $\hbar$ - The Planck constant.
 
 A lot of criticism have arisen \cite{SG, LW, HC1, SH, HM, MR, JL, YM} on Verlinde's assumption (1), perhaps due to an apparent contradiction with the dependence of the entropy on the area of a surface : $\Delta S \propto \Delta A$, where $\Delta A$ denotes the area variation. Is is also doubtful whether we can apply the HP for arbitrary surfaces or it is valid only for horizons. 
 
 He and Ma \cite{HM} had a tentative to compare the variation of the black hole (BH) entropy due to its radius variation when a test particle of mass $m$, located near the BH horizon, is moved with $\Delta x$. The relation they obtained seems to have no a reasonable physical meaning.
 
 A thorough analysis of the Verlinde conjecture (1) have been given by Modesto and Randono \cite{MR}. They noticed that the mass $m$ should lay close to the surface $\Sigma$ (the holographic screen, located between the source mass $M$ and the particle) at a distance of the order of the Compton wavelenght $\hbar/mc$. The change in entropy of $\Sigma$ is so given by eq. (1), being proportional to the change in radial distance of the test mass from $\Sigma$. Moreover, they identify the temperature  with Unruh's temperature of the causal horizon of a congruence of accelerated observers. They also remarked that there are two seemingly competing definitions of entropy ($\Delta S \propto \Delta A$ and $\Delta S \propto \Delta x$), where $\Delta x$ is taken as the distance from the horizon. 
 
 Gao \cite{SG} pointed out that the gravitational force cannot be an entropic force since gravity manifests even when $\Delta x = 0$. In addition, the Verlinde's covariant expression for $\Delta S$ introduces a time dependent entropy (from the 4-th component of $\partial S/\partial x^{a}$ ), even though one may have $\Delta x = 0$ \cite{HC1}. 
 
 J. Lee, Kim and S. Lee \cite{LKL} suggested that gravity is a quantum entanglement force connecting gravity to information using the Landauer principle. In their view, there is an energy $E_{h}$ related to information erasing at the horizon : $dE_{h} = T_{h} dS_{h}$, where $T_{h}$ is the horizon temperature and $dS_{h}$ is the horizon entropy change due to information erasing. However, their theory assumes neither the proportionality of entropy with the distance nor the entropic force. 
 
 Myung \cite{YM} even abandoned the linear relation (1) between $\Delta S$ and $\Delta x$ and works only with $\Delta S \propto \Delta A$. A relativistic quantum particle cannot be localized to better than its Compton wavelength and, therefore, it is indistinguisable from the horizon if the particle is away from the horizon (or the holographic screen) at a distance less than its Compton wavelength. 
 
 Recently Duncan, Myrzakulov and Singleton \cite{DMS} have studied the entropic derivation of the law of Newton for circular motion and showed that (1) has to be modified for the Verlinde model to work. In addition, Sakalli \cite{IS} highlighted the significance of the entropic force arising between a charged dilatonic black hole and a test particle, extending Verlinde's model to spacetimes with unusual asymptotics.
 
  In Verlinde's paper, $m$ seems to be a test particle, that is its gravitational field is neglected. In that case it is not clear what is the mechanism of the causal relation : displacement of $m$ - entropy increase. Much more realistic is, in our opinion, Fursaev's model \cite{DF} where the area of the screen changes under the influence of the gravitational field of $m$. A shift of the mass $m$ leads to a finite change of the screen area and, from here, of the entanglement entropy. \\

\section{ Conversion of rest energy in horizon energy}
 We will pay attention in this letter to the manner in which the test particle is moved with $\Delta x$. Since any change of the state of motion means aceleration, we take into account the simplest case : the particle is put in motion with constant acceleration $a$ (the hyperbolic motion) - a constant force acts on the particle, measured in the inertial rest-system). Therefore, a Rindler horizon will form, from the point of view of the accelerated observer, the ''distance to the horizon'' being given by $c^{2}/a$. In Minkowski coordinates the horizon is the light cone which is expanding with the speed of light ; the effect of the force applied to the particle is to stop the causal horizon at a fixed distance with respect to the accelerated observer. 
 
 When an increasing force acts on the mass $m$, $c^{2}/a$ shrinks and the Rindler horizon is located closer and closer to the particle (the location of the Rindler horizon at the distance $c^{2}/a$ has been recently remarked by Padmanabhan \cite {TP2}; he also analysed the case of a trajectory with time dependent acceleration $a(t)$ and found that the horizon is located at $c^{2}/a(t)$). We will adopt Bekenstein's hypothesis that, when $m$ approaches the horizon a distance less than its Compton wavelength, it merges with it (the fact that our horizon is not a BH horizon is of no importance, as we shall see).\\
  We consider that, once the particle is in the previous situation, its energy $mc^{2}$ is converted into the horizon energy $E_{h}$ \cite{HC1, JL} and this induces an increase of the horizon entropy $\Delta S_{h}$ 
\begin{equation}
mc^{2} = \Delta E_{h} = T_{U} \Delta S_{h}, 
\label{2}
\end{equation}
where $T_{U} = a \hbar/2 \pi c k_{B}$ is the Unruh temperature of the horizon.

 Since the mass $m$ travelled up to the horizon, we must have $\Delta x = c^{2}/a = \hbar/mc$ (see also \cite{MCP}), where $a$ is here the particle acceleration at the moment it approached the horizon at a distance $\hbar/mc$ (we notice that the above situation corresponds to $a \Delta x =  c^{2}$, namely, the potential $a \Delta x$ equals the maximum possible value). Hence, Eq. (2) yields
\begin{equation}
\Delta S_{h} = mc^{2} \frac{2 \pi c k_{B}}{\hbar} \frac{\Delta x}{c^{2}} = 2 \pi k_{B} \frac{mc}{\hbar} \Delta x.
\label{3}
\end{equation}
 In other words, the relation (1) is obtained, but when the following restrictions are imposed: $\Delta x$ to be equal to $c^{2}/a$ (the special value chosen by Lee in \cite{JL}) and the particle to merge with the horizon. Therefore, in our opinion, the Verlinde equation (1) is valid only in the case his holographic screen is a horizon \footnote{The authors of \cite{DMS} named the holographic screen in the case of linear acceleration a ''Rindler horizon''. However, Verlinde never used this name for his holographic screen. To my knowledge, the first time the name ''Rindler horizon'' has been given to the Verlinde holographic screen was in \cite{HC1}, pp.2} (Rindler, Hawking or cosmological) and $\Delta x$ is not arbitrary but equals ''the distance to the horizon'' $c^{2}/a$. Hence, (1) is a consequence of (2). 
 
 Another support for our option comes from the example below. Let us consider, following Verlinde, that we would shrink the screen to a radius $R_{0} < R$, with $\Delta R = R - R_{0} << R$. The entropy of the screen will be multiplied by $(R_{0}/R)^{2}$. If the particle $m$ stays at radius $R$ , the situation is equivalent with a displacement of it with $\Delta x = \Delta R$, away from the screen. \\
 The variation of the screen entropy is given by
 \begin{equation}
 \Delta S = S_{0} - S = - S \left[1 - (\frac{R_{0}}{R})^{2}\right] \approx - S \frac{2 R \Delta R}{R^{2}} = - \frac{2 \pi c^{3} k_{B} R \Delta R}{G \hbar},
 \label{4}
 \end{equation}
 where the expressions $S = k_{B} A/4 l_{P}^{2}$ and $S_{0} = k_{B} A_{0}/4 l_{P}^{2}$ have been used ($A = 4 \pi R^{2}$ and $A_{0} = 4 \pi R_{0}^{2}$). \\
 The variation of the entropy according to the formula (1) looks as 
 \begin{equation}
 \Delta S = - 2 \pi k_{B} \frac{mc}{\hbar} \Delta R.
 \label{5}
 \end{equation}
 A comparison between (4) and (5) leads to $R = Gm/c^{2}$, namely $R$ should be half the gravitational radius of $m$, a result which seems unphysical ($R$ and $m$ are independent parameters). This is an extra reason to abandon (1) as being valid for any $\Delta x$. 
 
 As far as the thermodynamical relation
 \begin{equation}
 F \Delta x = T \Delta S
 \label{6}
 \end{equation}
 is concerned, it should be written, in fact, in the form 
 \begin{equation}
 F = T \frac{\Delta S}{\Delta x}
 \label{7}
 \end{equation}
 which has a different interpretation compared to (6) \cite{YM}.
 
 Let us see now what is the role played by the mass $M$ which is surrounded by the Verlinde holographic screen. Outside the mass $M$, the geometry is Schwarzschild's. We know that, near a BH horizon, the geometry is that of Rindler. The mass $m$ being near the horizon, we may replace the Rindler horizon with the event horizon of a BH and the acceleration $a$ becomes the surface gravity \cite{HC2}, so that the Newton law of gravity is obtained .
 
 With the holographic screen viewed as a Rindler horizon, one may explain the origin of inertia. The accelerated observer comoving with the particle perceives the Rindler horizon behind him (if he accelerates) or in front of him (if he decelerates). Therefore, the spacetime beyond the horizon continually disappears behind the horizon \cite{JL}. Applying the Landauer principle here, we find that this information erasing demands the energy consumption $\Delta E_{h} = T_{U} \Delta S_{h}$. Consequently, to ''drag'' the horizon, a force given by $F \Delta x = \Delta E_{h}$ will arise (for a Minkowski observer there is no force because the horizon is not dragged - it is expanding with the speed of light).\\

 \section{Time dependent entropy}
  Verlinde \cite{EV} has written a covariant expression for $\Delta S$ (taking his $a$ and $b$ to run from 0 to 3). Suppose $N_{a}$ has a nonzero temporal component $N_{t} < 0$. We have in this case
\begin{equation}
\frac{\partial S}{\partial t} = - 2\pi k_{B} \frac{mc^{2}}{\hbar} N_{t} 
\label{8}
\end{equation}
If the mass $m$ is at rest with respect to the screen ( $\Delta x = 0$), $S$ depends only on time and (1) can be written as 
\begin{equation}
\Delta S = 2 \pi k_{B} \frac{mc^{2}}{\hbar} N_{t} \Delta t .
\label{9}
\end{equation}

Eq. (9) shows that we have an entropy change simply because time flows. That is in accordance with Gao's remark \cite{SG} that the condition $\Delta x \neq 0$ is not mandatory to get a nonzero gravitational force. If the entropy increases on the holographic screen as time proceeds, that means Nature does work for that and the corresponding energy is recovered on the screen. 

We could show the above argument works without any test particle $m$. Let us consider an observer at rest with respect to an inertial system in flat space. After a time $\Delta t$, his causal horizon expands , covering a sphere of radius $c \Delta t$. Because of the new informations acquired by our observer \cite{HC3}, an entropy variation $\Delta S$ given by 
\begin{equation}
\Delta S = \frac{1}{2} k_{B}~ \Delta N 
\label{10}
\end{equation}
will arise, localized on the causal horizon, considered as a holographic screen. $\Delta N$ above is given by
\begin{equation}
\Delta N = \alpha ~\frac{A}{l_{P}^{2}}
\label{11}
\end{equation}
where $\alpha$ is a constant of the order of unity, $A = 4 \pi (c \Delta t)^{2}$ is the area of the causal horizon after $\Delta t$ and $l_{P} = (G \hbar/c^{3})^{1/2}$ . $\Delta S$ from (10) leads to an energy variation on the screen, given by $\Delta E = T \Delta S$. To get the temperature $T$ we make use of the Heisenberg Principle , applied for the energy per degree of freedom $\epsilon(T) \equiv (1/2) k_{B} T$ 
\begin{equation}
\epsilon(T) ~\Delta t = \beta \hbar
\label{12}
\end{equation}
where $\beta$ is another constant of the order of unity. With $T$ from (12) the energy change becomes
\begin{equation}
\Delta E = 4 \pi \alpha \beta~ \frac{c^{4}}{G}~ c~ \Delta t
\label{13}
\end{equation}
i.e $\Delta E$ is proportional to the radius $c \Delta t$ of the sphere (or to the time elapsed from an arbitrary origin). This resembles the dependence of the black hole mass on the horizon radius : $M_{bh} = (c^{2}/2G) R_{H}$. Relying on this analogy, we choose $\alpha = 1/4$ and $\beta = 1/2 \pi$. \\
Dividing Eq. (13) by $\Delta r = c~ \Delta t$, one obtains $\Delta E/\Delta r \equiv F =  c^{4}/2G$, a value akin with that obtained by Easson et al. \cite{EFS} in their study on a cosmological entropic force (see also \cite{CLL, WLW}).

 We could formally define a ''surface gravity'' $\kappa$ on the screen, by analogy with the black hole case (see \cite{AV} for a generalisation of the surface gravity)
 \begin{equation}
 \kappa = \frac{c^{4}}{4 G M} = \frac{c}{2 \Delta t}
 \label{14}
 \end{equation}
 It is worth mentioning that the observer appears to be inside the holographic screen which, in addition, is going away with the velocity of light (see also \cite{HC4} for a model of the black hole interior). A possible explanation of the nature of $\Delta E$ , appearing even in Minkowski spacetime, has been given in \cite{HC5}.
  We see that all the above physical quantities are time dependent (we should have used, for example, $\Delta T$ instead of $T$). Because of the simplicity of the relations, we think the model works as if we had thermodynamics at equilibrium. That explains why the black hole temperature is not time dependent : its event horizon (which acts as the causal horizon) is not expanding (the light emitted from the horizon cannot escape outside). Therefore, the Hawking temperature could be obtained from Eq. (12) replacing $c \Delta t$ by $4 R_{H}$. In other words, $T$ is time independent when the causal horizon (a null surface) is not expanding.\\

 \section{Acceleration and the entropy gradient} 
  Verlinde considers the particle with mass $m$ approaches the screen , when it should merge with the degrees of freedom of the screen. The number of bits $\Delta N$ carried by the particle follows from
 \begin{equation}
 mc^{2} = \frac{1}{2} k_{B} T \Delta N,
 \label{15}
 \end{equation}
  whence he immediately obtained
\begin{equation}
\Delta S = \frac{1}{2} k_{B} \frac{a \Delta x}{c^{2}} \Delta N.
\label{16}
\end{equation}
What is $\Delta x$ here? Since the particle merged with the microscopic degrees of freedom on the screen, we cannot have in (16) an arbitrary $\Delta x$. Note that Verlinde used the Unruh formula for the temperature $T$. But Unruh's thermal bath comes from the fact that the hyperbolic (uniformly accelerated) observer has a horizon (the screen plays the role of a local Rindler horizon). Therefore, $\Delta x$ should be $c^{2}/a$, the distance to the horizon (the special role played by this value was also remarked in \cite{JL, HC6}) . Hence, Eq. (9) yields
\begin{equation}
\Delta S = \frac{1}{2} k_{B} \Delta N ,
\label{17}
\end{equation}
which is in accordance with Gao's estimation \cite{SG}
\begin{equation}
mc^{2} = \Delta E = T \Delta S = \frac{1}{2} k_{B} T \Delta N .
\label{18}
\end{equation}
 It is worth mentioning that in the Unruh expression for $T$, $a$ is not a vector, as Verlinde has suggested, but the modulus of the acceleration 4-vector $a^{b}$. The Unruh formula is obtained using quantum and special - relativistic arguments and has no a classical counterpart. 
 
 Take, for example, the Rindler metric
 \begin{equation}
 ds^{2} = - c^{2}(1-gx/c^{2})^{2} dt^{2} + dx^{2} + dy^{2} + dz^{2} ,
 \label{19}
 \end{equation}
 where $g$ is the rest-system acceleration and the horizon is located at $x = c^{2}/g$. \\
 An $x = const.$ (static) observer will have the 4 - acceleration
 \begin{equation}
 a^{b} = (0, - \frac{g}{1 - gx/c^{2}}, 0, 0) ,
 \label{20}
 \end{equation}
 with the modulus $a = (a^{b} a_{b})^{1/2} = g/(1 - gx/c^{2})$. It equals $|a^{x}|$ because $a^{b}$ has only one nonzero component (in fact, $a$ is the acceleration from the equation for the hyperbolic trajectory, $x^{2} - c^{2}t^{2} = (c^{2}/a)^{2})$. 

 We see that $a$ is $x$ - dependent and is equal to $g$ at the origin of coordinates (i.e., far from the horizon). The surface gravity on the horizon can be obtained from
 \begin{equation}
 \kappa = \sqrt{a^{b} a_{b}} ~\sqrt{- g_{tt}} |_{x = c^{2}/g} = g 
 \label{21}
 \end{equation}
 and, as in the case of the surface gravity ($ = c^{4}/4GM$ ) for a Schwarzschild black hole, it is measured from infinity, which is equivalent to ''far from the horizon''. \\
 Therefore, using $\ddot{x}$ or $a^{x}$ in the Unruh formula is not appropriate. A similar opinion has recently been expressed by Cai et al. \cite{CCO}. 

 Let us consider, as an example, the acceleration a proton must possess for its Rindler horizon to lie at a distance of the order of the associated Compton wavelength. One obtains
  \begin{equation}
  a = \frac{mc^{3}}{\hbar} \approx 10^{34} \frac{cm}{s^{2}}.
  \label{22}
  \end{equation}
  This corresponds to a temperature of $10^{13} K$ or $1 ~GeV$ ($m$ is here the proton mass and the temperature has been obtained from $mc^{2} = 2 \pi k_{B} T$). Such a huge acceleration we measure during an experiment of RHIC (Relativistic Heavy Ions Collisions). The incoming particle travels with a speed close to the speed of light and is suddenly damping (when the collision takes place). The accelerated proton sees the vacuum as a physical medium of temperature $T_{U}$ \cite{CKS}.\\
  
  \section{Conclusions}  
   We have argued that one could convert the rest energy of a particle into horizon energy when its acceleration is such that the Rindler horizon is located at a distance from the particle less than its associated Compton wavelength.\\
Since the relations $\Delta S \propto \Delta A$ and $\Delta S \propto \Delta x$ seem to be in contradiction, we found that the second expression is not valid for any $\Delta x$ but only when it equals ''the distance to the horizon'' $c^{2}/a$. From the covariant form of $\Delta S$ and by means of the Uncertainty Principle we suggest that the energy per degree of freedom on the holographic screen and its temperature undergo quantum fluctuations.

The conversion rest energy - horizon energy could be checked in a high acceleration regime arising during ultrarelativistic collisions. \\

\textbf{References} \\

\end{document}